\documentclass[pre,twocolumn,showpacs,preprintnumbers,amsmath,amssymb]{revtex4-1}

\usepackage{graphicx}
\usepackage{dcolumn}
\usepackage{bm}

\begin{document}

\newcommand{\bea}{\begin{eqnarray}}
\newcommand{\eea}{  \end{eqnarray}}
\newcommand{\bit}{\begin{itemize}}
\newcommand{\eit}{  \end{itemize}}

\newcommand{\be}{\begin{equation}}
\newcommand{\ee}{\end{equation}}
\newcommand{\ra}{\rangle}
\newcommand{\la}{\langle}
\newcommand{\U}{\widetilde{U}}


\def\bra#1{{\langle#1|}}
\def\ket#1{{|#1\rangle}}
\def\bracket#1#2{{\langle#1|#2\rangle}}
\def\inner#1#2{{\langle#1|#2\rangle}}
\def\expect#1{{\langle#1\rangle}}
\def\e{{\rm e}}
\def\proj{{\hat{\cal P}}}
\def\tr{{\rm Tr}}
\def\H{{\hat H}}
\def\Hdag{{\hat H}^\dagger}
\def\Lop{{\cal L}}
\def\Ehat{{\hat E}}
\def\Edag{{\hat E}^\dagger}
\def\Shat{\hat{S}}
\def\Sdag{{\hat S}^\dagger}
\def\Ahat{{\hat A}}
\def\Adag{{\hat A}^\dagger}
\def\U{{\hat U}}
\def\Udag{{\hat U}^\dagger}
\def\Zhat{{\hat Z}}
\def\Phat{{\hat P}}
\def\Op{{\hat O}}
\def\id{{\hat I}}
\def\x{{\hat x}}
\def\P{{\hat P}}
\def\Px{\proj_x}
\def\Pr{\proj_{R}}
\def\Pl{\proj_{L}}


\title{Quantum Isoperiodic Stable Structures and Directed Transport}

\author{Gabriel G. Carlo}
\affiliation{Departamento de F\'\i sica, CNEA, Libertador 8250, (C1429BNP) Buenos Aires, Argentina}
\email{carlo@tandar.cnea.gov.ar}

\date{\today}


\begin{abstract}

It has been recently found that the so called isoperiodic stable structures (ISSs) have a 
fundamental role in the classical current behavior of dissipative ratchets 
[Phys. Rev. Lett. {\bf 106}, 234101 (2011)]. Here I analyze their quantum counterparts, 
the quantum ISSs (QISSs), which have a fundamental role in the quantum current behavior. 
QISSs have the simple attractor shape of those ISSs which settle down 
in short times. However, in the majority of the cases they are strongly different from the ISSs, 
looking approximately the same as the quantum chaotic attractors that are at their vicinity 
in parameter space. By adding thermal fluctuations of the size of $\hbar_{\rm eff}$ to the 
ISSs I am able to obtain very good approximations to the QISSs. I conjecture that in general, 
quantum chaotic attractors could be well approximated by means of just the classical information 
of a neighboring ISS plus thermal fluctuations. I expect to find this behavior in 
quantum dissipative systems in general.

\end{abstract}

\maketitle

During the last years there have been great advances in the 
area of directed transport \cite{Feynman,Reimann,Kohler}. 
Understood as transport phenomena in periodic systems out of equilibrium,
this field has attracted great attention giving rise to 
many investigations of interdisciplinary nature. In fact, ratchet models have found 
application in biology \cite{biology}, nanotechnology \cite{nanodevices}, and chemistry 
\cite{chemistry}, just to name a few examples. In particular, cold atoms in optical lattices have shown 
very successful theoretical developments and implementations \cite{CAexp,AOKR}. 
Moreover, Bose-Einstein condensates have been transported 
by means of the so called purely quantum ratchet accelerators \cite{BECratchets}, where 
the current has no classical counterpart \cite{purelyQR} and 
the energy grows ballistically \cite{recentStudies,coherentControl}.

The current generation mechanism consists of breaking all spatiotemporal symmetries 
leading to momentum inversion \cite{origin}. Classical deterministic ratchets with dissipation 
are generally associated with an asymmetric chaotic attractor \cite{Mateos}.
Quantum effects were considered to analyze the first so-called quantum ratchets in \cite{Qeffects},
while a dissipative quantum ratchet, interesting for cold atoms experiments 
has been introduced in \cite{qdisratchets}. Very recently, the parameter space of the classical 
counterpart of this system has been studied in detail \cite{Manchein}. There it has been 
found that not only chaotic domains but more importantly, a family of isoperiodic stable 
structures (ISSs) has a fundamental role in understanding the current behavior, a 
major issue in any ratchet system. Moreover, a complete characterization of this family 
has been given and a connection with the current values has been also provided. 

Then, it is natural to ask how this family of ISSs translates into the quantum domain. 
In this letter I answer this question. I have found that the classical decay time towards these 
stable structures is a determining factor for the shape of the corresponding quantum ISSs (QISSs). 
The majority of these classical structures have very long transient times making the QISSs look 
like the quantum chaotic attractors at their vicinity in parameter space. 
In comparatively few cases I have found quantum structures similar to these classically 
simple objects (periodic points in the case of maps). On the other hand, by adding thermal 
fluctuations of the size of $\hbar_{\rm eff}$ to the classical system very good approximations 
to the QISSs were obtained. 

I investigate a paradigmatic dissipative ratchet system given by the map \cite{qdisratchets,Manchein}
\begin{equation}
\left\{
\begin{array}{l}
\overline{n}=\gamma n + 
k[\sin(x)+a\sin(2x+\phi)],
\\
\overline{x}=x+ \tau \overline{n},
\end{array}
\right.
\label{dissmap} 
\end{equation}
where $n$ is the momentum variable conjugated to $x$, $\tau$ is the period of the map 
and $\gamma$ is the dissipation parameter. 
This dynamics can be interpreted as that of a particle moving in one dimension 
[$x\in(-\infty,+\infty)$] in a periodic kicked asymmetric potential:
\begin{equation}
V(x,t)=k\left[\cos(x)+\frac{a}{2}\cos(2x+\phi)\right]
\sum_{m=-\infty}^{+\infty}\delta(t-m \tau),
\end{equation}
where $\tau$ is the kicking period, and subject to a dissipation $0\le \gamma \le 1$.
When $\gamma=0$ the particle is overdamped and for $\gamma=1$ we recover the 
conservative evolution. As usual, the directed transport 
appears due to broken spatial ($a \neq 0$ and $\phi \neq m \pi$) and temporal ($\gamma \neq 1$) 
symmetries. It is useful to notice that the classical dynamics only depends on the 
parameter $K=k \tau$ by means of introducing the rescaled momentum $p=\tau n$.

In order to quantize this model I follow the standard procedure: 
$x\to \hat{x}$, $n\to \hat{n}=-i (d/dx)$ ($\hbar=1$).
Since $[\hat{x},\hat{p}]=i \tau$, the effective Planck constant 
is $\hbar_{\rm eff}=\tau$. The classical limit corresponds to 
$\hbar_{\rm eff}\to 0$, while $K=\hbar_{\rm eff} k$ remains constant.  
The final ingredient, the dissipation, is introduced by means of the 
master equation \cite{Lindblad} for the density operator $\hat{\rho}$ of the 
system 
\begin{equation}
\dot{\hat{\rho}} = -i 
[\hat{H}_s,\hat{\rho}] - \frac{1}{2} \sum_{\mu=1}^2 
\{\hat{L}_{\mu}^{\dag} \hat{L}_{\mu},\hat{\rho}\}+
\sum_{\mu=1}^2 \hat{L}_{\mu} \hat{\rho} \hat{L}_{\mu}^{\dag}. 
\label{lindblad}
\end{equation}
Here $\hat{H}_s=\hat{n}^2/2+V(\hat{x},t)$ is the system
Hamiltonian, \{\,,\,\} is the anticommutator, and $\hat{L}_{\mu}$ are the Lindblad operators 
given by 
\begin{equation}
\begin{array}{l}
\hat{L}_1 = g \sum_n \sqrt{n+1} \; |n \rangle \, \langle n+1|,\\
\hat{L}_2 = g \sum_n \sqrt{n+1} \; |-n \rangle \, \langle -n-1|,
\end{array}
\end{equation} 
with $n=0,1,...$ and $g=\sqrt{-\ln \gamma}$ (due to the Ehrenfest theorem). 
In all cases I have evolved $10^6$ classical random initial conditions 
having $p \in [-\pi,\pi]$ and $x \in [0,2\pi]$ ($<p_0>=0$), and their quantum 
counterpart.

\begin{figure}[htp]
\vspace{1.0cm}
\includegraphics[angle=0.0, width=8.0cm]{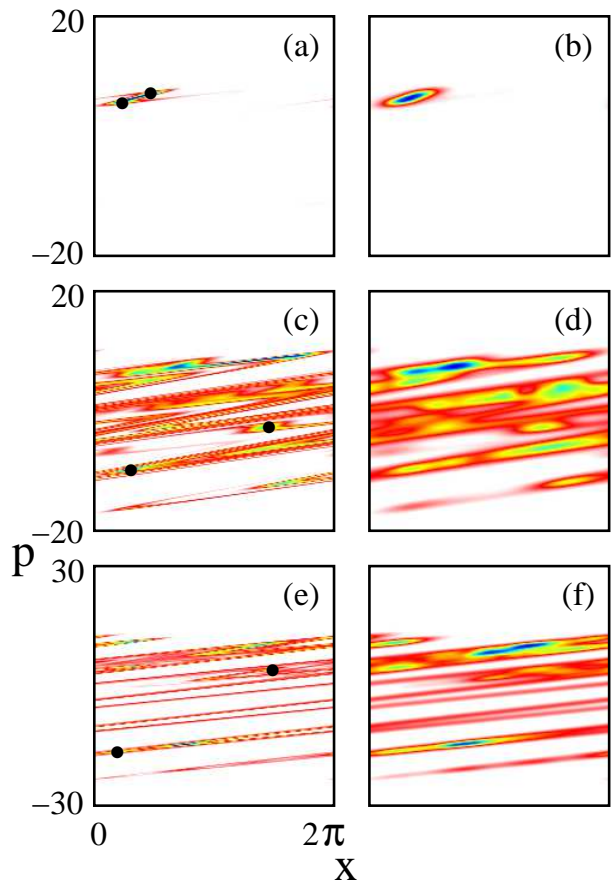}
\caption{(Color online) Phase space distributions corresponding to the ISSs labeled by $B_1$ 
($\gamma=0.2$, $k=8.2$, first row), $C_{-1}$ ($\gamma=0.64$, $k=5.6$, middle row), 
and $D_{-1}$ ($\gamma=0.29$, $k=11.9$, bottom row) in \cite{Manchein}. Lower to higher values 
of these distributions go from white to (rainbow colors) grays.
Left column shows the classical Poincar\'e maps while the right column 
shows the Husimi functions. In (a), (c), and (e) the periodic points of 
each ISS are marked by means of a black dot. The surrounding distributions 
correspond to the ISSs at finite $T$ (for (a) $T=0.058$, for (c) $T=0.12$, and 
for (e) $T=0.07$). In (b), (d), and (f) $\hbar_{\rm eff} \simeq 0.082$. Notice that for (e) and (f) 
$p \in [-30,30]$, while for the rest $p \in [-20,20]$.}
\label{fig1}
\end{figure}

In \cite{Manchein} the important role that ISSs have on the ratchet currents 
corresponding to dissipative systems has been shown. There, three main kinds 
were identified and called $B_M$, $C_M$ and $D_M$, where $M$ stands for an 
integer or rational number and corresponds to the mean momentum of these 
structures in units of $2\pi$. With perhaps the only exception of $\gamma 
\rightarrow 1$ (i.e., near the conservative limit), ISSs organize the 
parameter space structure and then are essential to understand the current 
behavior. Though I have checked the general validity of the 
results by sampling several points in the parameter space, 
I have selected a representative case for each kind of these structures. Also, 
I have studied a chaotic attractor case in the vicinity of 
one of them, for comparison purposes. In Fig. \ref{fig1}. I show the phase 
space portraits (after $50$ time steps of the map) corresponding to the chosen cases, i.e. $B_1$ 
($\gamma=0.2$, $k=8.2$, Figs. \ref{fig1} (a) and (b)), $C_{-1}$ 
($\gamma=0.64$, $k=5.6$, Figs. \ref{fig1} (c) and (d)), and $D_{-1}$ 
($\gamma=0.29$, $k=11.9$, Figs. \ref{fig1} (c) and (d)). Only for the first 
case I have found a point like structure (besides the uncertainty restrictions) 
similar to the classical simple attractors marked with black dots in Figs. \ref{fig1} (a), 
(c), and (e). The two other kinds (and even other regions of 
the same $B_1$ large structure in parameter space, and also other $B_M$ with $M \neq 1$) 
behave like a quantum chaotic attractor. I have found that 
this surprising behavior is due to chaos induced by quantization. 
The way to prove this is to introduce fluctuations of the 
size of $\hbar_{\rm eff}$ in the classical model so as to induce classical chaos.
I do this by changing $\overline{n}' \rightarrow \overline{n}$ in 
Eq. \ref{dissmap}, where $\overline{n}'=\overline{n} + \xi$. The thermal 
noise $\xi$ is related to $\gamma$, according to $ <\xi^2> =2 (1-\gamma) k_B T$, 
where $k_B$ is the Boltzmann constant (which I take equal to 1) and $T$ 
is the temperature. Finally I take $T \simeq \hbar_{\rm eff}/[2 (1-\gamma)]$. 
As a result, I obtain the classical distributions that surround the 
periodic points and that are shown in Figs. \ref{fig1} (a), (c), and (e). 
They are remarkably similar to the corresponding quantum chaotic attractors. 
Even in the first case (i.e. Figs. \ref{fig1} (a) and (b) corresponding to 
$B_1$), adding fluctuations is sufficient to reproduce the quantum simple attractor 
(which is not a point, of course). 
This is confirmed by means of the overlap measure defined as 
$O=\iint {\cal L}(x,p) {\cal H}(x,p) dx dp$ (${\cal L}(x,p)$ and ${\cal H}(x,p)$ 
are normalized phase space distributions with the same discretization) which compares 
the whole distributions, not just the first moment. In this case these distributions correspond 
to the classical Liouville distribution and the quantum Husimi function. 
For $B_1$ and $T=0$ I obtain $O \simeq 0.1$, while for $T=0.058$ the result 
is $O \simeq 0.87$, the other two cases behave the same way, resulting in 
$O \simeq 0.07$ and $O \simeq 0.91$ for $T=0$ and $T=0.12$ respectively in the 
$C_{-1}$ case, and $O \simeq 0.14$ and $O \simeq 0.87$ for $T=0$ and $T=0.07$ 
respectively in the $D_{-1}$ ISS. 
 
%
%
%
%

\begin{figure}[htp]
\hspace{0.0cm}
\includegraphics[angle=0.0, width=8.0cm]{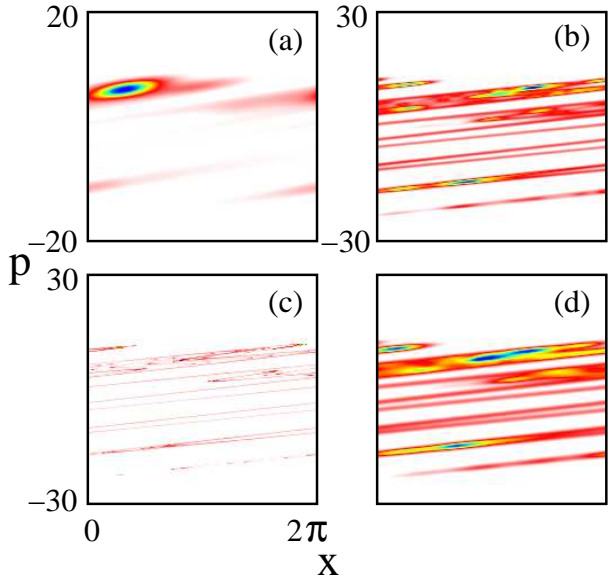}
\caption{(Color online) In (a) I show the Husimi function for the $B_1$ QISS (see Figs. \ref{fig1}(a) and (b)) 
for $\hbar_{\rm eff} \simeq 0.246$. The same (color) gray scale as in Fig \ref{fig1} is used. 
In (b) the $D_{-1}$ QISS for $\hbar_{\rm eff} \simeq 0.027$ is shown. In (c) and (d), 
the classical ($T=0$) and quantum ($\hbar_{\rm eff} \simeq 0.082$) 
chaotic attractors respectively, neighboring $D_{-1}$, for which $\gamma=0.26$ and $k=11.9$.}
\label{fig2}
\end{figure}

I have found that the convergence to a quantum simple attractor is not uniform with 
respect to $\hbar_{\rm eff}$, but highly dependent on the ISS in question. To illustrate this result 
I show the QISS corresponding to the $B_1$ case for $\hbar_{\rm eff} \simeq 0.246$ in Fig. \ref{fig2}(a). 
It is clear that for this higher value of $\hbar_{\rm eff}$ the QISS starts to resemble 
the neighboring chaotic attractor. On the contrary, when I study the QISS corresponding to $D_{-1}$ 
for $\hbar_{\rm eff} \simeq 0.027$ (see Fig. \ref{fig2}(b)) the chaotic nature of the quantum 
attractor shows no sign of vanishing. On the other hand, it is also very important to point out 
that the QISS have great influence on the quantum chaotic attractors surrounding them. 
In fact, I have taken $\gamma=0.26$ and $k=11.9$ in the chaotic area $A$ (according to \cite{Manchein} 
nomenclature), in the vicinity of the ``shrimp`` structure $D_{-1}$. Here the 
classical attractor resembles the quantum one without the need of additional fluctuations, 
giving an overlap  $O \simeq 0.73$ at $T=0$. In this case the chaotic mixing is 
enough to recover the quantum chaotic shape. But, most importantly, the quantum chaotic 
attractor is very similar to the corresponding QISS, their overlap (between the distributions 
shown in Figs. \ref{fig1}(f) and \ref{fig2}(d)) being $O \simeq 0.99$. This suggests 
the idea that general quantum chaotic attractors could be acceptably reproduced with just the classical 
information corresponding to a neighboring ISS plus thermal fluctuations. This gives to the QISSs an 
even more relevant role in the understanding of quantum dissipative ratchets and also of 
quantum dissipative systems in general.
%
%
%
%

\begin{figure}[htp]
\hspace{0.0cm}
\includegraphics[angle=0.0, width=8.0cm]{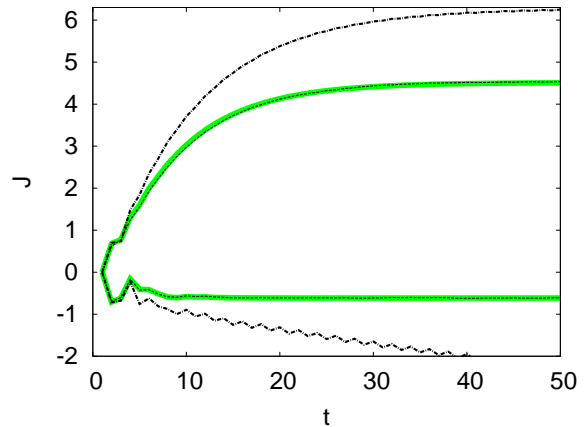}
\caption{(Color online) Classical and quantum current $J$ as a function of map time steps 
$t$. Black dot-dashed lines correspond to classical values at $T=0$, 
black dashed thin lines to classical values at the finite temperatures of 
Fig \ref{fig1}, and (green) gray full lines to the quantum values. 
Upper lines correspond to the $B_1$ ISS, and lower ones to $C_{-1}$.}
\label{fig3}
\end{figure}

\begin{figure}[htp]
\hspace{0.0cm}
\includegraphics[angle=0.0, width=8.0cm]{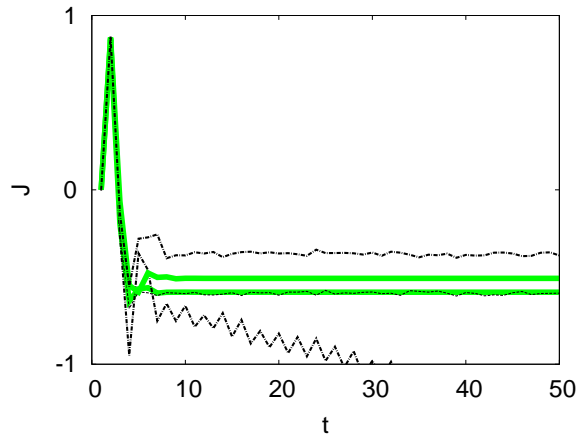}
\caption{(Color online) Same as in Fig \ref{fig3}, but for the $D_{-1}$ ISS (lower lines) and 
the chaotic attractor in its vicinity shown in Fig. \ref{fig2} (c) and (d) (upper lines).}
\label{fig4}
\end{figure}

The influence of QISSs and the effect of thermal noise in 
the ratchet current $J=<p>$ (where $<>$ stands for either the classical or quantum averages) 
can be seen in Figs. \ref{fig3} and \ref{fig4}, where I compare the classical and 
quantum $J$ for the cases of Fig. \ref{fig1} and the chaotic attractor of Figs. \ref{fig2}(c) and (d). 
The upper lines of Fig. \ref{fig3} represent the $B_1$ case. The dot-dashed black curve corresponds to the 
classical current at $T=0$, which stabilizes in a relatively short time ($50$ time steps) 
when compared to the same curve for the $C_{-1}$ case in the same figure 
(in fact, this latter one stabilizes in more than $250$ time steps). 
In general, the ISSs have long transients, settling down in times one order of magnitude 
longer than this. This seems to be the main reason for the $B_1$ case 
under scrutiny to show a quantum simple behavior in contrast with the other cases. I am currently 
investigating the details of this behavior \cite{future}.

In this Figure, the (green) gray full lines correspond to the quantum current, while the black dashed 
thin lines correspond to the classical $J$ at finite $T$. The effects of the thermal environment 
quickly stabilize this current. This is specially evident in the $C_{-1}$ case, which clearly loses even the 
period $2$ bumps exhibited by the lower dot-dashed black line. 
Moreover, in both cases displayed in Fig. \ref{fig3} the agreement of the classical $J$ at finite $T$ with 
the quantum current is excellent, reflecting what I have already shown by means of the phase space distributions.
%
%
%
%

In Fig. \ref{fig4} I make the same comparison as in Fig. \ref{fig3}, but in this case for 
the QISS corresponding to the $D_{-1}$ kind with the chaotic attractor in its vicinity. 
Again, the classical current for the ISS stabilizes very slowly (it takes around $700$ time 
steps). On the other hand, the quantum $J$ stabilizes very quickly (in $7 \sim 8$ time steps) as does 
the classical one at finite $T$ (their agreement being excellent). We can see that the classical 
current corresponding to the chaotic attractor case also stabilizes very quickly, without the need 
of additional fluctuations. In the classically chaotic case the stabilization of the current and its 
main features (like the generic low values and lack of periodic fluctuations) are already determined 
by the chaotic mixing.
%
%
%
%

In conclusion, I have found that QISSs have a fundamental role in the behavior 
of the current in quantum dissipative ratchets, as ISSs have in their classical counterparts. This 
has been unveiled by means of analyzing a paradigmatic system in the ratchet and 
chaos (classical and quantum) literature. QISSs have the simple attractor shape of the 
classical ISSs only in the few cases where the time in which they stabilize is very short. 
But in general they have an extended, chaotic attractor shape in phase space. 
I have found that the behavior of the QISSs can 
be understood by means of just the classical information contained in the corresponding ISSs plus 
thermal fluctuations of the order of $\hbar_{\rm eff}$. Moreover, the quantum chaotic attractors 
in their vicinity have a very similar structure. This makes us conjecture that in general 
it should be possible to approximate any quantum chaotic attractor by means of the essential 
classical information contained in a neighboring ISS (plus stochastic fluctuations). 
This is expected to be a general result valid for any quantum dissipative system which 
has a contractive (in phase space) kind of noise and it will be the matter of future investigations \cite{future}.

\vspace{3pc}

Financial support form CONICET is gratefully acknowledged. I would like to thank fruitful discussions 
with C. Manchein and M. Beims. 


\end{document}